\documentclass[aps,superscriptaddress,eqsecnum,nofootinbib,showpacs,preprintnumbers]{revtex4}
\usepackage{graphicx,epsfig}
\usepackage{amsmath}
\usepackage {amssymb}

\newcommand{\be}{\begin{eqnarray}}
\newcommand{\ee}{\end{eqnarray}}
\newcommand{\bea}{\begin{eqnarray}}
\newcommand{\nn}{\nonumber}
\newcommand{\eea}{\end{eqnarray}}

\def\k{\kappa}
\def\m{\mu}
\def\n{\nu}

\def\s{\sigma}

\def\k{\kappa}
\def\m{\mu}
\def\n{\nu}

\def\s{\sigma}

\newcommand{\fr}{\frac}

\begin{document}

\title{Superradiant Amplification of a Scalar Wave Coupled Kinematically to Curvature Scattered off   a  Reissner-Nordstr\"om Black Hole }

\author{Theodoros Kolyvaris}
\email{teokolyv@mail.ntua.gr} \affiliation{Instituto de
F\'{i}sica, Pontificia Universidad Cat\'olica de Valpara\'{i}so,
Casilla 4059, Valpara\'{i}so, Chile.}

\author{Eleftherios Papantonopoulos}
\email{lpapa@central.ntua.gr} \affiliation{Physics Division,
National Technical University of Athens, 15780 Zografou Campus,
Athens, Greece.}

\date{\today}

\begin{abstract}

We study the dynamics of a massive charged scalar wave which except its canonical coupling to gravity it is also coupled kinematically to Einstein tensor. We find that as the strength of the new coupling  is increased the scattered wave off the horizon of a Reissner-Nordstr\"om black hole is  superradiantly amplified resulting to the
instability
of the Reissner-Nordstr\"om spacetime.

\end{abstract}

\maketitle

\section{Introduction}

It was shown by Penrose that if a particle accretes  into a Kerr black hole then this process may result to the extraction of energy from the black hole.
 This process suggests that energy is extracted
from the black hole. The same process can be realized if you scatter waves off the black hole. It was found by Misner that this happens if
the relation
\be \omega < m \Omega \label{rad1} \ee
is satisfied, where $
\omega$ is the frequency of the incitant wave and $\Omega$ is the rotational frequency of the black hole. Later it was shown by
Teukolsky  that this
applification mechanism also works in the case of electromagnetic and gravitational waves if the condition (\ref{rad1}) is satisfied.
Further Bekenstein showed \cite{Bekenstein} that the Misner process can also be realized by
extracting charge and electrical energy from a
charged black hole. He showed that this is a  consequence of the
Hawking's theorem that the surface area of a black
hole cannot decrease.

Introducing a reflecting mirror it was suggested  by   Press and Teukolsky \cite{PressTeu1} that the wave will be amplified
as a result of the
bounce back and forth between the black hole and the mirror. In the case of a rotating black hole, the role of the reflecting mirror can be played by the mass of the scalar field because the  gravitational force binds the massive field and
keeps it from escaping to infinity once the condition    (\ref{rad1}) is satisfied.
 As a consequence, the rotational energy
extracted from the black hole by the incident field grows
exponentially over time. A detailed investigation of the superradiant amplification of a wave scattered off a Kerr black hole surrounded by a mirror,  was carried out in \cite{Cardoso:2004nk}.
In the case of extracting charge and electric energy from a  Reissner-Nordstr\"om black hole, the Misner condition (\ref{rad1}) is modified \cite{Bekenstein}  to
\begin{equation}\label{rad2}
\omega<q\Phi ~,
\end{equation}
where $q$ is the charge coupling constant of the field and
$\Phi$ is the electric potential of the charged black hole.
The superradiant scattering
of charged scalar waves in the regime (\ref{rad2}) may lead to an
instability of the Reissner-Nordstr\"om spacetime.

The stability of Reissner-Nordstr\"om  black holes under  neutral gravitational
and electromagnetic perturbations was established  by
Moncrief \cite{Monf1,Monf2}. Evidence was provided in \cite{Hod:2013eea,Hod:2013nn,Hod:2015hza,Hod:2016kpm}  for the
stability of charged Reissner-Nordstr\"om  black holes  under charged scalar
perturbations.  The stability of
extremal braneworld charged  holes was studied in \cite{Zhang:2013haa} were it was shown that if the spacetime dimension is higher than four,
the superradiant amplification can occur.
 Also  the Einstein-Maxwell-Klein-Gordon equations
for a spherically symmetric scalar field
scattering off a Reissner-Nordstr\"om  black hole in asymptotically flat spacetime were considered in \cite{Baake:2016oku} and a superradiant instability was found. The superradiance instability of  charged black holes placed in a cavity was studied in \cite{Herdeiro:2013pia,Degollado:2013bha,Sanchis-Gual:2015lje,Sanchis-Gual:2016tcm} (for a recent review on superradiance see \cite{Brito:2015oca}).

The recent developments in AdS/CFT correspondence \cite{Maldacena:1997re} introduces an AdS spacetime  as a natural reflecting boundary on which the
reflecting wave keeps amplified  driving the system unstable. This results in the formation of hairy  black holes with a charged scalar field trapped outside the black hole in which the electric repulsion balances the scalar condensate against gravitational collapse. Hairy black holes were constructed in  global AdS$_5$ spacetime in \cite{Basu:2010uz,Bhattacharyya:2010yg,Dias:2011tj,Basu:2016mol}. The superradiant amplification of a wave packet may result to the destabilization of the AdS spacetime itself \cite{Cardoso:2004hs,Cardoso:2013pza,Bosch:2016vcp,Dias:2016pma}. If we perturb the AdS spacetime with a scalar field  the system evolves towards the formation of a black hole \cite{Bizon:2011gg,Dias:2012tq,Buchel:2012uh,Bizon:2015pfa}. The stability of near  extremal and extremal  charged hairy black hole solution under charged massive scalar field perturbations was studied in \cite{Gonzalez:2017shu}.

The application of the  AdS/CFT correspondence to condensed matter
systems (for a review see \cite{Hartnoll:2009sz}) had revived the
interest on the dynamics of a scalar field outside a black hole
horizon. The transition of metalic state to a superconducting
state which is a strongly-coupled problem in condensed matter
physics can be described by its dual weekly-coupled gravity
problem using the AdS/CFT correspondence \cite{Maldacena:1997re}.
The simplest holographic superconductor model
\cite{Hartnoll:2008vx,Hartnoll:2008kx}
 is described by an Einstein-Maxwell-scalar
field theory in a planar AdS spacetime while an
exact gravity dual of a holographic superconductor was discussed
in \cite{Koutsoumbas:2009pa}. The dynamics of a holographic superconductor depends crucially on
the behaviour of a scalar field near the horizon of a black hole. A mechanism of trapping the scalar field
just outside the horizon of a charged black hole was presented in \cite{Gubser:2005ih,Gubser:2008px}. It was shown that the effective mass of the
scalar field becomes negative  breaking in this
way an Abelian gauge symmetry outside the horizon of the
Reissner-Nordstr\"om black hole resulting to an instability of the Reissner-Nordstr\"om spacetime.

In the case of de-Sitter  charged black holes instabilities were found  in higher dimensions. It was showed in \cite{Konoplya:2008au,Konoplya:2013sba}
 that higher dimensional Reissner-Nordstr\"om-de Sitter black holes are gravitationally unstable for large values of the electric charge in $D≥7$ spacetime dimensions. The existence of
such instability was proved analytically in the near-extremal limit \cite{Cardoso:2010rz}. In four dimensions
a superradiance instability was found \cite{Zhu:2014sya} in the  Reissner-Nordstr¨om-de Sitter black holes
against charged scalar perturbations with vanishing angular momentum, l = 0.

In this work we study the effect of charged scalar perturbations on the stability of the Reissner-Nordstr\"om black hole of a scalar field which except its canonical coupling to gravity it is also coupled kinematically  to Einstein tensor.
This term belongs to a general class of
scalar-tensor gravity theories resulting from the Horndeski
Lagrangian \cite{Horndeski:1974wa}. These theories, which were
recently rediscovered \cite{Deffayet:2011gz},
 give second-order field equations and contain as a subset a theory
which preserves classical Galilean symmetry \cite{Nicolis:2008in,
Deffayet:2009wt, Deffayet:2009mn}.

 The derivative coupling of the scalar field to Einstein tensor
introduces a new scale in the theory which on short distances allows
to find  black hole solutions \cite{Kolyvaris:2011fk,Rinaldi:2012vy,Kolyvaris:2013zfa,Babichev:2013cya},
while if one considers the gravitational collapse of a scalar field coupled to the Einstein tensor then a  black hole is formed
 \cite{Koutsoumbas:2015ekk}.
On large distances  the presence of the derivative coupling acts as a friction term in the inflationary period of the cosmological
evolution  \cite{Amendola:1993uh,Sushkov:2009hk,germani}. Moreover, it was found that at the end of
inflation in the preheating period, there is a suppression of heavy particle production  as the derivative coupling is increased. This was attributed to the fast decrease of  kinetic
energy of the scalar field due to its  wild oscillations \cite{Koutsoumbas:2013boa}. This change of the kinetic energy of the scalar field to Einstein tensor allowed to holographically simulate the effects of a high concentration of impurities in a material \cite{Kuang:2016edj}.

The above discussion indicates that one of the main effects of the kinematic coupling of a scalar field to Einstein tensor
is that  gravity  influences strongly  the  propagation of the scalar field compared to a scalar field minimally coupled to gravity.
We will use this behaviour of the charged scalar field coupled to Einstein tensor to study its behaviour outside the horizon of a Reissner-Nordstr\"om black hole.   Will  show
that this new dimensionful
derivative coupling  provides a scale for a confining potential, which is an effect
similar to the AdS radius provided by the cosmological constant, and in the same time modifies the Bekenstein's superradiance condition (\ref{rad2}). Then for a wide range of parameters satisfying the supperadiant condition, the charged scalar field will be trapped in this confining potential leading to a superradiant instability of the Reissner-Nordstr\"om black hole. The stability of  Reissner-Nordstr\"om and Kerr spacetimes   was discussed in \cite{Chen:2010qf,Chen:2010ru,Ding:2010fh} in a different context, calculating the quasinormal frequencies and  the greybody factors  in the presence of the derivative coupling.

The  work is organized as follows. In Section \ref{instability} we will review the stability of a Reissner-Nordstr\"om black hole under charge scalar perturbations of a scalar field minimally coupled to gravity. In Section \ref{dercoupl} we introduce the derivative coupling of a scalar field to the Einstein tensor and we discuss the stability of the background Reissner-Nordstr\"om black hole under charge scalar field perturbations.

\section{Supperradiant Instability}

\label{instability}

In this section we will review the superradiant stability of the Reissner-Nordstr\"om black hole discussed in \cite{Hod:2013eea,Hod:2013nn}. Consider a Reissner-Nordstr\"om black hole of mass $M$ and
electric charge $Q$ with a metric
\begin{equation}\label{Eq3}
ds^2=-f(r)dt^2+{1\over{f(r)}}dr^2+r^2(d\theta^2+\sin^2\theta
d\phi^2)\ ,
\end{equation}
where
\begin{equation}\label{Eq4}
f(r)\equiv 1-{{2M}\over{r}}+{{Q^2}\over{r^2}}\  .
\end{equation}

A massive charged scalar field was considered in the Reissner-Nordstr\"om spacetime and its dynamics was described by the
 the Klein-Gordon equation
\begin{equation}\label{Eq5}
[(\nabla^\nu-iqA^\nu)(\nabla_{\nu}-iqA_{\nu}) -\mu^2]\Psi=0\  ,
\end{equation}
where $A_{\nu}=-\delta_{\nu}^{0}{Q/r}$ is the electromagnetic
potential of the black hole. Here $q$ and $\mu$ are the charge and
mass of the field, respectively. Consider a scalar wave of the form
\begin{equation}\label{Eq6}
\Psi_{lm}(t,r,\theta,\phi)=e^{im\phi}S_{lm}(\theta)R_{lm}(r)e^{-i\omega
t}\ ,
\end{equation}
where $\omega$ is the conserved frequency of the mode, $l$ is the
spherical harmonic index, and $m$ is the azimuthal harmonic index
with $-l\leq m\leq l$. We expect an instability for $\omega_I>0$
\cite{Kokkotas:1999bd,Nollert:1999ji}.
 In the above decomposition $R$ and $S$ denote the
radial and angular part and the radial Klein-Gordon equation is given by \cite{Hod:2013nn}
\begin{equation}\label{Eq7}
\Delta{{d} \over{dr}}\Big(\Delta{{dR}\over{dr}}\Big)+UR=0\ ,
\end{equation}
where
\begin{equation}\label{Eq8}
\Delta\equiv r^2-2Mr+Q^2\  ,
\end{equation}
and
\begin{equation}\label{Eq9}
U\equiv(\omega r^2-qQr)^2 -\Delta[\mu^2r^2+l(l+1)]\  .
\end{equation}

To solve the radial equation (\ref{Eq7}) we impose boundary conditions of purely ingoing waves at the
black-hole horizon and a decaying  solution at spatial infinity
\begin{equation}\label{Eq11}
R \sim e^{-i (\omega-qQ/r_H)y}\ \ \text{ as }\ r\rightarrow r_H\ \
(y\rightarrow -\infty)\ ,
\end{equation}
and
\begin{equation}\label{Eq12}
R \sim y^{-iqQ}e^{-\sqrt{\mu^2-\omega^2}y}\ \ \text{ as }\
r\rightarrow\infty\ \ (y\rightarrow \infty)\  ,
\end{equation}
where the ``tortoise" radial coordinate $y$ is defined by
$dy=(r^2/\Delta)dr$. For frequencies in the superradiant regime
(\ref{rad2}), the boundary condition (\ref{Eq11}) describes an
outgoing flux of energy and charge from the charged black hole
\cite{Bekenstein,PressTeu1}. As it was shown in \cite{Damour:1976kh} these boundary conditions
lead to  a
discrete set of resonances $\{\omega_n\}$ which correspond to the
bound states of the charged massive field.

Defining a new $\psi$ function
\begin{equation}\label{Eq14}
\psi\equiv \Delta^{1/2}R\  ,
\end{equation}
the equation (\ref{Eq7}) can be written in
the form of a Schr\"odinger-like wave equation
\begin{equation}\label{Eq15}
{{d^2\psi}\over{dr^2}}+(\omega^2-V)\psi=0\  ,
\end{equation}
where
\begin{equation}\label{Eq16}
\omega^2-V={{U+M^2-Q^2}\over{\Delta^2}}\  .
\end{equation}

Then it was shown in \cite{Hod:2013eea,Hod:2013nn} that there is no superradiant amplification  for the charged Reissner-Nordstr\"om black holes because mainly two necessary conditions  cannot be satisfied
simultaneously namely, the condition of  superradiant amplification (\ref{rad2}) and the existence of a trapping potential well.

\section{Superrrandiance Instability with the Derivative Coupling}
\label{dercoupl}

In this section  we  introduce the kinematic coupling of the scalar field to Einstein tensor. The action we consider reads
\be \label{action}
S_0=\int dx^{4} \sqrt{-g}\; \left[\fr{R}{16\pi G}-(g^{\m\n}-\kappa G_{\m\n})D^{\m}\Psi(D^{\n}\Psi)^*-\frac{1}{4}F^{\m\n}F_{\m\n}-\m^2|\Psi|^2\right]~,
\ee
where $D_\m\equiv \nabla_\m-iqA_\m$ and $F_{\mu\nu} = \partial_\mu A_\nu - \partial_\nu A_\mu$. In what follows $G = c = 1$.
Varying the action (\ref{action}) we get the Einstein equations
\bea G_{\mu\nu} = 8\pi T_{\mu\nu} \
\ , \ \ \ \ T_{\mu\nu} = T_{\mu\nu}^{(\varphi)} +
T_{\mu\nu}^{(EM)} - \kappa\Theta_{\mu\nu}~, \label{einst}
\eea
where $T_{\mu\nu}^{(\varphi)}$,
$T_{\mu\nu}^{(EM)}$  are the energy-momentum tensors of the scalar and electromagnetic fields
\bea
T_{\mu\nu}^{(\varphi)} & = &   \Psi_{\mu\nu} + \Psi_{\nu\mu} - g_{\mu\nu}(g^{ab}\Psi_{ab} + m^2 |\varphi|^2)~, \\
T_{\mu\nu}^{(EM)} & = & F_{\mu}^{\phantom{\mu} \alpha} F_{\nu
\alpha} - \fr{1}{4} g_{\mu\nu} F_{\alpha\beta}F^{\alpha\beta}~,
\eea
while  $\Theta_{\mu\nu}$ is the contribution to the energy-momentum tensor of the $G_{\mu \nu}$ term
\bea
\Theta_{\mu\nu}  = & -& g_{\mu\nu} R^{ab}\Psi_{ab} + R_{\nu}^{\phantom{\nu}a}(\Psi_{\mu a} + \Psi_{a\mu}) + R_{\mu}^{\phantom{\mu}a} (\Psi_{a\nu} + \Psi_{\nu a})  - \fr{1}{2} R (\Psi_{\mu\nu} + \Psi_{\nu\mu}) \nn\\
& - & G_{\mu\nu}\Psi - \fr{1}{2}\nabla^a\nabla_\mu(\Psi_{a\nu} + \Psi_{\nu a}) - \fr{1}{2}\nabla^a\nabla_\nu(\Psi_{\mu a} + \Psi_{a\mu}) + \fr{1}{2}\Box (\Psi_{\mu\nu} + \Psi_{\mu\nu}) \nn \\
& + & \fr{1}{2}g_{\mu\nu} \nabla_a\nabla_b (\Psi^{ab} + \Psi^{ba})
+ \fr{1}{2}(\nabla_\mu\nabla_\nu + \nabla_\nu\nabla_\mu) \Psi -
g_{\mu\nu}\Box\Psi~. \label{theta} \eea
The Klein-Gordon equation is
\be
(\partial\mu-i q A_\mu) \left[
\sqrt{-g}(g^{\mu\nu} - \kappa G^{\mu\nu})(\partial\nu - i q A_\nu)\Psi \right] =
\sqrt{-g} \m^2 \Psi~, \label{glgord}
\ee
and the Maxwell equations are
\be
\nabla_\nu F^{\mu\nu} +(g^{\mu\nu} - \kappa G^{\mu\nu}) \left[ 2 q^2 A_\nu |\Psi|^2 + i q
(\Psi^*\nabla_\nu\Psi - \Psi\nabla_\nu\Psi^*)\right]=0~.
\label{max}
\ee

In the background of a Reissner-Nordstr\"om black  metric
\bea
ds^2 = -f(r)dt^2 + \fr{dr^2}{f(r)} + r^2 d\Omega^2~,
\eea
with $f(r) = 1 - 2M/r + Q^2/r^2$ and an electromagnetic potential $A_t = -\frac{Q}{r}$
we decompose the scalar field as in (\ref{Eq6})
\bea
\Psi = e^{-i\omega t}e^{i m \varphi}S(\theta)\frac{\psi(r)}{r}~. \label{wav11}
\eea
Then the radial equation resulting from the Klein-Gordon equation using (\ref{wav11}) can be written in a Schr\"odinger-like form
\bea\label{Schr}
\frac{d^2\psi}{dr_*^2}+U_{eff}(r)\psi(r_*)=0~,
\eea
where we have defined the ``tortoise" coordinate
\bea
\frac{dr_*}{dr} = \frac{r^2}{f(r) \left(r^2+\kappa(1-f-rf')\right)}
\eea
and the effective potential is given by
\bea
U_{eff} &=& \left(\omega -\frac{q Q}{r}\right)^2-f \left[\frac{\ell(\ell+1)}{r^2}+\mu ^2+\frac{f'}{r^2}\right] \nonumber\\
&+&\kappa  \left(-\frac{(f-1) \left(2 f^2-f \ell(\ell+1)+2 q^2 Q^2\right)}{r^4}+\frac{\left(f \mu ^2-2 \omega ^2\right) f'}{r}\right.\nonumber\\
&+&\left.\frac{4 (f-1) q Q \omega +2 \left(f \left[f-1+\ell(\ell+1)\right]-q^2 Q^2\right) f'}{r^3} \right.\nonumber\\
&+&\left.\frac{2 (f-1) \left(f \mu ^2-2 \omega ^2\right)+8 q Q \omega  f'+4 f \left(f'\right)^2+f \left[2 f+\ell(\ell+1)\right]
 f''}{2 r^2}\right)\nonumber\\
&+& \kappa ^2 \left(\frac{(f-1)^2 \left(2 f^2+q^2 Q^2\right)}{r^6}+\frac{\omega ^2 \left(f'\right)^2}{r^2}\right.\nonumber\\
&+&\left.\frac{(f-1) \left(-2 (f-1) q Q \omega +\left(f (1+f-\ell(\ell+1))+2 q^2 Q^2\right) f'\right)}{r^5}\right.\nonumber\\
&-&\left.\frac{f' \left(-4 (f-1) \omega ^2+4 q Q \omega  f'+2 f \left(f'\right)^2+f \left(2 f+\ell(\ell+1)\right) f''\right)}{2 r^3}
\right.\nonumber\\
&+&\left.\frac{(f-1) \left(2 (f-1) \omega ^2-f \left(2 f+\ell(\ell+1)\right) f''\right)}{2 r^4}\right.\nonumber\\
&-&\left.\frac{8 (f-1) q Q \omega  f'+2 \left(f \left(2(f-1)+\ell(\ell+1)\right)-q^2 Q^2\right) \left(f'\right)^2}{2 r^4}\right)~. \label{Effpot}
\eea

Note that the effective potential because of the presence of the derivative coupling $\kappa$ has the lapse function at second order and this  introduces high order terms in the radial coordinate $r$. Then the effective potential depends on seven parameters $U_{eff}(M, Q, \mu, q, \omega, l, \kappa)$ and even its numerical study is difficult.

The effective potential at infinity goes like
\bea
U_\infty \sim \omega^2-\m^2~,
\eea
while near the horizon can be approximated as
\bea
U_{r_H} \sim \frac{(r_H^2+\kappa \Phi^2)^2(\omega-q\Phi)^2}{r_H^4} +O(r-r_H)~,
\eea
where $\Phi=Q/r_H$ is the electric potential at the horizon.

In a scattering experiment the Klein-Gordon equation \eqref{Schr} has the following asymptotic behaviour
\bea
\psi\sim \left\{ \begin{array}{ll}
Te^{-i\sigma r_*} & \textrm{, as } r\rightarrow r_H\\
e^{-ir_*\sqrt{\omega^2-\m^2}} + Re^{ir_*\sqrt{\omega^2-\m^2}} & \textrm{, as } r\rightarrow\infty
\end{array} \right.
\eea
where we have set
\bea
\sigma \equiv \frac{(r_H^2+\kappa \Phi^2)(\omega-q\Phi)}{r_H^2}~.
\eea

These boundary conditions correspond to an incident wave of unit amplitude from spatial infinity giving rise to a reflected wave of amplitude $R$ and a transmitted wave of amplitude $T$ at the horizon.

Since the effective potential is real, there exists another solution $\bar{\psi}$ to \eqref{Schr} which satisfies the complex conjugate boundary conditions \cite{Ching:1993gt}. The solutions $\psi$ and $\bar{\psi}$ are linearly independent and thus their Wronskian $W=\psi\frac{d}{dr_*}\bar{\psi}-\bar{\psi}\frac{d}{dr_*}\psi$ is independent of $r_*$.
Evaluating the Wronskian at the horizon and infinity respectively, we get
\bea
W(r\rightarrow r_H) &=& 2 i \sigma |T|^2, \\
W(r\rightarrow \infty) &=& -2 i (|R|^2-1)\sqrt{\omega^2-\mu^2}
\eea
and by equating the two values we get,
\bea
|R|^2 &=& 1 - \frac{\s}{\sqrt{\omega^2-\mu^2}}|T|^2~.
\eea
We can see that if $\s < 0$ the wave is superradiantly amplified, $|R|^2 > 1$ \cite{Benone:2015bst}. So, the superradiant   condition (\ref{rad2}) in the presence of the derivative coupling is modified to
\bea\label{supercond}
(r_H^2+\kappa \Phi^2) (\omega - q \Phi) < 0~.
\eea

To see whether the superradiance will cause the instability of the Reissner-Nordstr\"om black hole, we need to check whether there exists a potential well outside the horizon to trap the reflected wave. If the potential well exists, the superradiant instability will occur and the wave will grow exponentially over time near the black hole to make the background Reissner-Nordstr\"om   black hole unstable.
We are interested in solutions of the radial equation \eqref{Schr} with the physical boundary conditions of purely ingoing waves
at the horizon and a decaying solution at spatial infinity. A bound state decaying exponentially at spatial infinity is characterized by $\omega^2 < \mu^2$. We choose the parameters $M, Q, \mu, q, \kappa, \omega$ to meet this condition together with the superradiant condition \eqref{supercond}. In what follows we will set $\ell=0$ and fix the horizon at $r_H=1$.

\subsection{Case I: $\omega<q\Phi$ and $r_H^2+\kappa\Phi^2>0$}

We will first examine the case where $\omega<q\Phi$ and $\kappa>0$ so that \eqref{supercond} is satisfied. Our aim is to see if a trapping potential is formed outside the horizon of the black hole. Since the effective potential (\ref{Effpot}) is a high order polynomial in the radial coordinate $r$ and therefore its analytical treatment  is difficult, we will relay on its numerical investigation.  We will fix the parameters $q$ and $\omega$,  $\mu$ as the relation $\omega^2 < \mu^2$ to be satisfied, and we will make a systematic study of the effective potential (\ref{Effpot}) varying the parameters $\kappa$ and $Q$, $M$. We will look for various values of the ratio  $Q/M$ and specially for values approaching one (near extremal case) and one (extremal limit).

\begin{figure}
\centering
\includegraphics[scale=1]{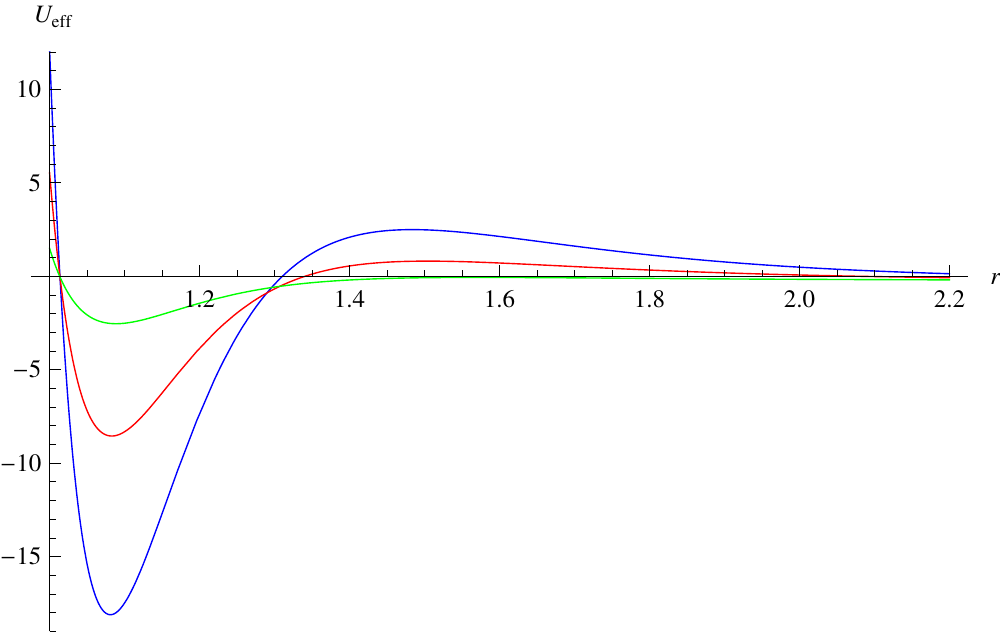}
\caption{The potential as a function of the radial coordinate for coupling constant $\kappa=150,\ \kappa=100,\ \kappa=50$ (blue, red, green), mass and charge for the black hole and the scalar field $M=0.625,\ Q=0.5,\ \mu=0.63,\ q=0.86$ respectively and $\omega=0.34$.} \label{fig1a}
\end{figure}

\begin{figure}
\centering
\includegraphics[scale=1]{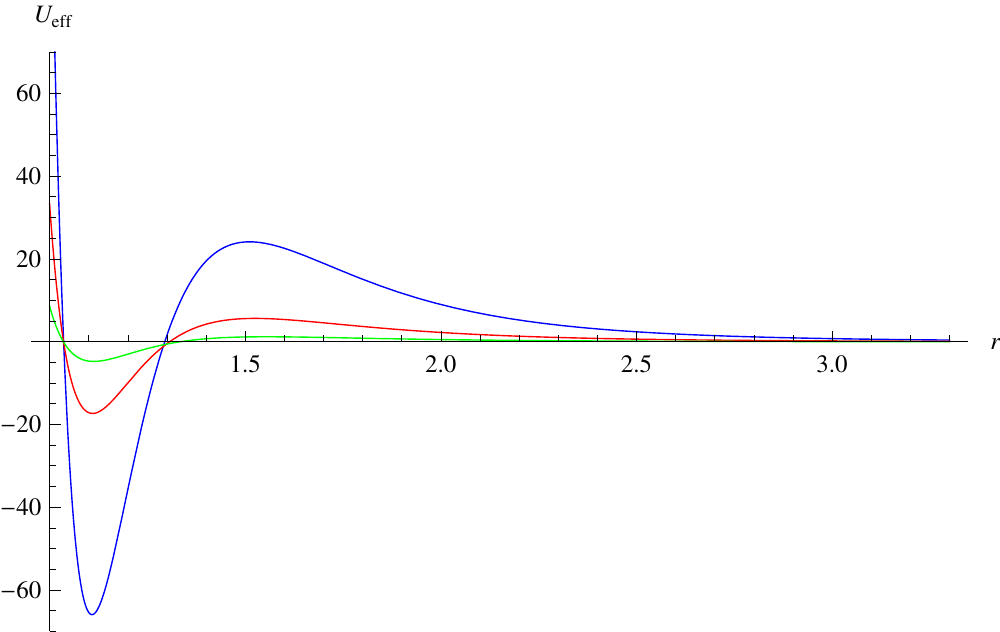}
\caption{The potential as a function of the radial coordinate for coupling constant $\kappa=200,\ \kappa=100,\ \kappa=50$ (blue, red, green), mass and charge for the black hole and the scalar field $M=0.82,\ Q=0.8,\ \mu=0.63,\ q=0.86$ respectively and $\omega=0.6$.} \label{fig1b}
\end{figure}

In Table \ref{table1} we give the ratio $(Q/M)^2$ for various values of $Q$ and $M$. For a characteristic value of $Q$ and $M$ in the first part of the Table \ref{table1} we depict the effective potential as a function of $r$ in  Figure \ref{fig1a}, while in Figure \ref{fig1b} we depict the effective potential as a function of $r$   for a characteristic value of the second part of the Table \ref{table1}. We observe that irrespective of the value of the ratio $Q/M$ a trapped potential is formed outside the horizon of the black hole and most importantly  as the derivative coupling  $\kappa$ is increased the potential well gets deeper. This suggests that as the strength of the coupling of the scalar to curvature is increased,  the superradiant instability is amplified.

\begin{figure}
\centering
\includegraphics[scale=1]{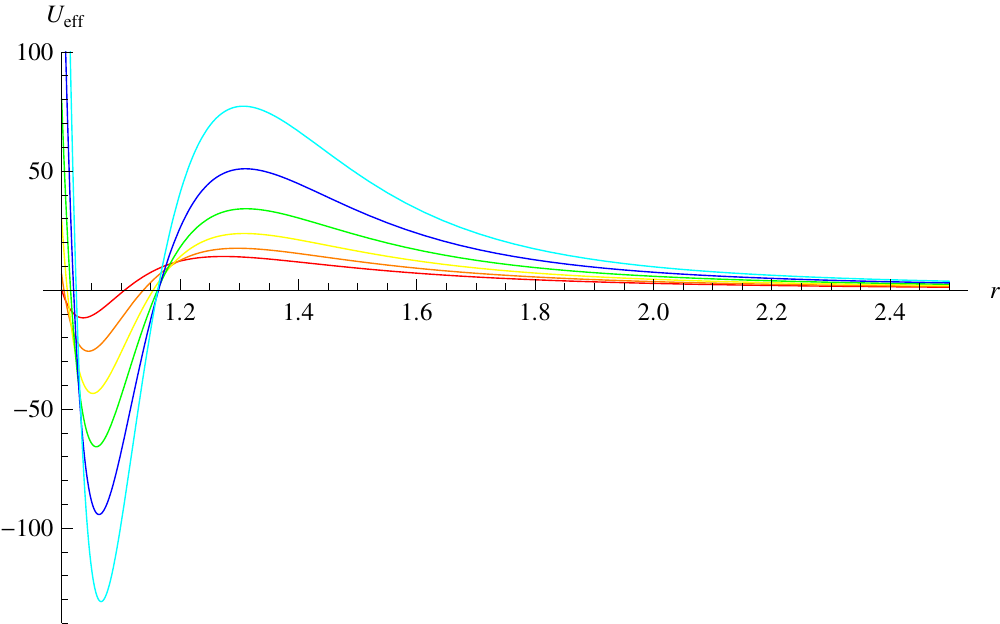}
\caption{The potential as a function of the radial coordinate for coupling constant $\kappa=10$, mass and charge for the scalar field $q=0.86,\ \mu=1.63,\ \omega=1.28$. The ratio $(Q/M)^2$ is 0.852, 0.808, 0.760, 0.720, 0.679, 0.64 (red, orange, yellow, green, blue, cyan) respectively.} \label{fig1c}
\end{figure}

Finally in Figure \ref{fig1c} we depict the effective potential as a function of $r$ for fixed values of $\omega, \mu, q, \kappa$ and for different values of $Q$ and $M$ belonging to the third part of Table \ref{table1}. We observe   that the potential well is deeper as the ratio $Q/M$ gets smaller. This corresponds to larger values of the mass $M$ and charge $Q$ since we are in the third part of  the Table \ref{table1}. This suggests that a highly charged massive black hole radiates more.


\begin{table}[h]
\begin{center}
\begin{tabular}{| c | c | c |}
\hline
\qquad M \qquad & \qquad Q \qquad &\qquad $(Q/M)^2$\qquad \\
\hline
 0.505 & 0.1 & 0.039 \\
 0.52 & 0.2 & 0.148 \\
 0.545 & 0.3 & 0.303 \\
 0.58 & 0.4 & 0.476 \\
 0.625 & 0.5 & 0.640 \\
 0.68 & 0.6 & 0.779 \\
 0.745 & 0.7 & 0.883 \\
 \hline
 0.82 & 0.8 & 0.952 \\
 0.905 & 0.9 & 0.989 \\
 1. & 1. & 1.000 \\
 1.105 & 1.1 & 0.991 \\
 1.22 & 1.2 & 0.967 \\
 1.345 & 1.3 & 0.934 \\
 1.48 & 1.4 & 0.895 \\
 \hline
 1.625 & 1.5 & 0.852 \\
 1.78 & 1.6 & 0.808 \\
 1.945 & 1.7 & 0.764 \\
 2.12 & 1.8 & 0.721 \\
 2.305 & 1.9 & 0.679 \\
 2.5 & 2. & 0.640\\
 \hline
\end{tabular}
\end{center}
\caption{The first and third sections correspond to a ratio less than 8/9, while the second greater than 8/9.}\label{table1}
\end{table}

\subsection{Case II: $\omega>q\Phi$ and $r_H^2+\kappa\Phi^2<0$}
We consider now the case where $\omega>q\Phi$ and $\kappa < - (r_H/\Phi)^2$ so again \eqref{supercond} is satisfied. This case is interesting.
 In \cite{Kolyvaris:2011fk, Kolyvaris:2013zfa}  fully backreacted black hole solutions were found in the presence of the derivative coupling $\kappa$. However, these solutions exists only in the case of positive coupling while if the coupling constant $\kappa$ is negative, then the system of Einstein-Maxwell-Klein-Gordon equations is unstable and no solutions were found.
 The sign of the coupling constant $\k$ was also discussed in \cite{Germani:2010gm}, where it was claimed that a negative $\kappa$ introduces ghosts into the theory. In \cite{Koutsoumbas:2013boa} a very small window of negative $\kappa$ was shown to be allowed. Finally in \cite{Kuang:2016edj} it was found that a negative $\kappa$ in a theory of holographic superconductors generates an instability in the gravity sector of the theory and using the gauge/gravity duality this was interpreted  as generating an impurity on a material on the boundary of the theory.

  In the present study the scalar field does not backreact on the metric. However, if for a range of parameters  relation \eqref{supercond}, for $\omega>q\Phi$ and a negative $\kappa$ is satisfied and there is a confining potential, then there should be an instability in the system of a charged scalar field scattered off a Reissner-Nordstr\"om black hole and this is due to superrandiant amplification.

\begin{figure}
\centering
\includegraphics[scale=1]{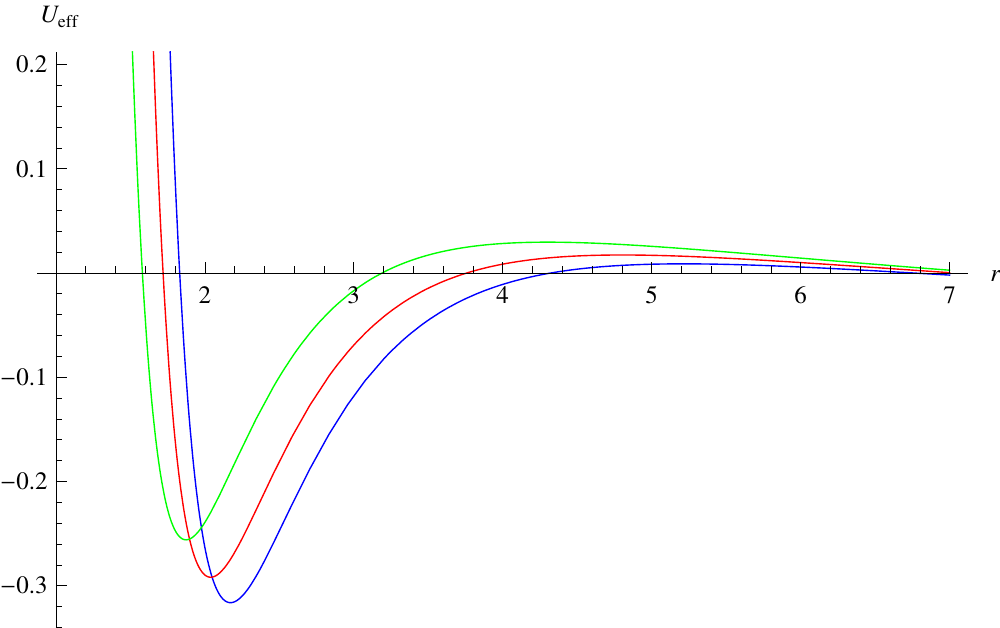}
\caption{The potential as a function of the radial coordinate for coupling constant $\kappa=-45,\ \kappa=-35,\ \kappa=-25$ (blue, red, green), mass and charge for the black hole and the scalar field $M=0.625,\ Q=0.5,\ \mu=1.63,\ q=1.6$ respectively and $\omega=1.6$.} \label{fig2a}
\end{figure}

\begin{figure}
\centering
\includegraphics[scale=1]{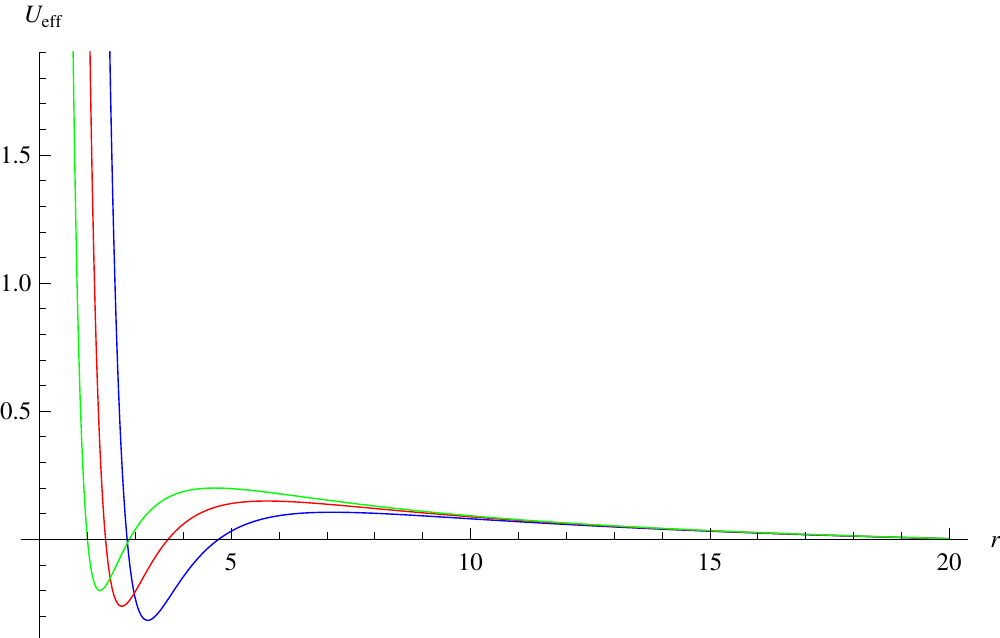}
\caption{The potential as a function of the radial coordinate for coupling constant $\kappa=-100,\ \kappa=-50,\ \kappa=-25$ (blue, red, green), mass and charge for the black hole and the scalar field $M=0.82,\ Q=0.8,\ \mu=1.63,\ q=0.9$ respectively and $\omega=1.6$.} \label{fig2b}
\end{figure}

 In Figures \ref{fig2a} and \ref{fig2b} we depict the effective potential as a function of $r$, again for two characteristic values of the ratio $Q$ and $M$ belonging to the first and second part of Table \ref{table1}.  As before, in both cases we found that there exists a negative minima of the potential well which gets deeper for larger (absolute) values of the derivative coupling $\kappa$.

\begin{figure}[h]
\centering
\includegraphics[scale=1]{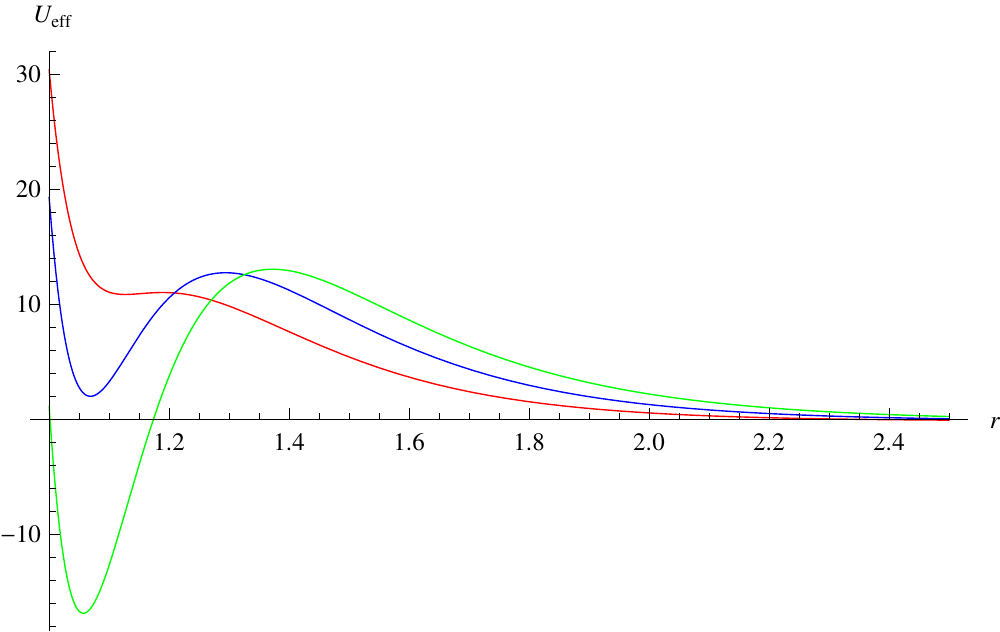}
\caption{The potential as a function of the radial coordinate for coupling constant $\kappa=-120$, mass and charge for the scalar field $q=0.86,\ \mu=0.63,\ \omega=0.62$. The ratio $(Q/M)^2$ is 0.64, 0.779, 0.883 (red, blue, green) respectively.} \label{fig2c}
\end{figure}

\begin{figure}[h]
\centering
\includegraphics[scale=1]{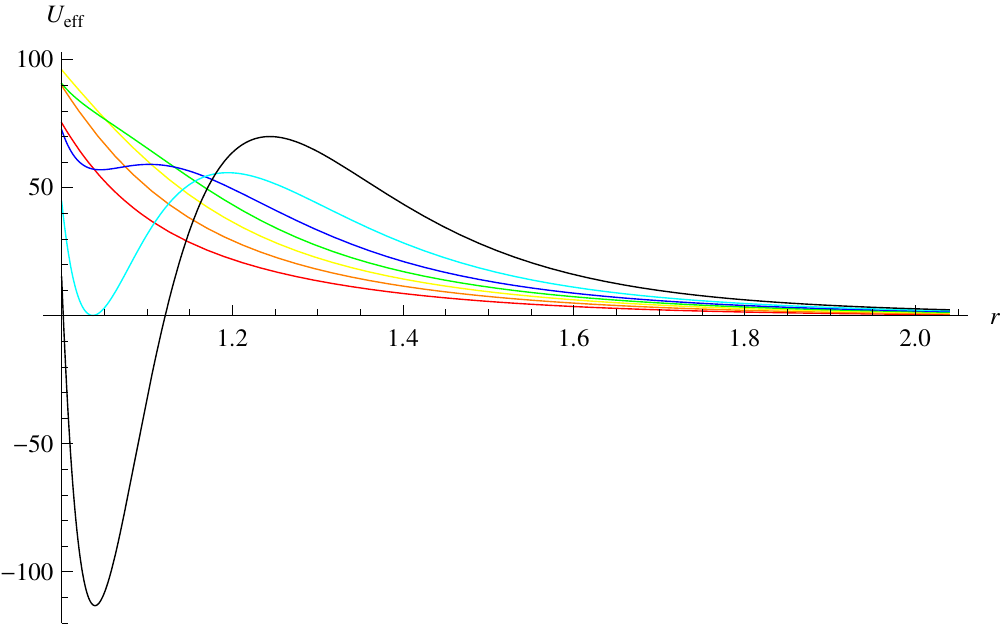}
\caption{The potential as a function of the radial coordinate for coupling constant $\kappa=-50$, mass and charge for the scalar field $q=0.4,\ \mu=0.63,\ \omega=0.6$. The ratio $(Q/M)^2$ is 0.952, 0.989, 1.000, 0.991, 0.967, 0.939, 0.895 (red, orange, yellow, green, blue, cyan, black) respectively.} \label{fig2d}
\end{figure}

\begin{figure}[h]
\centering
\includegraphics[scale=1]{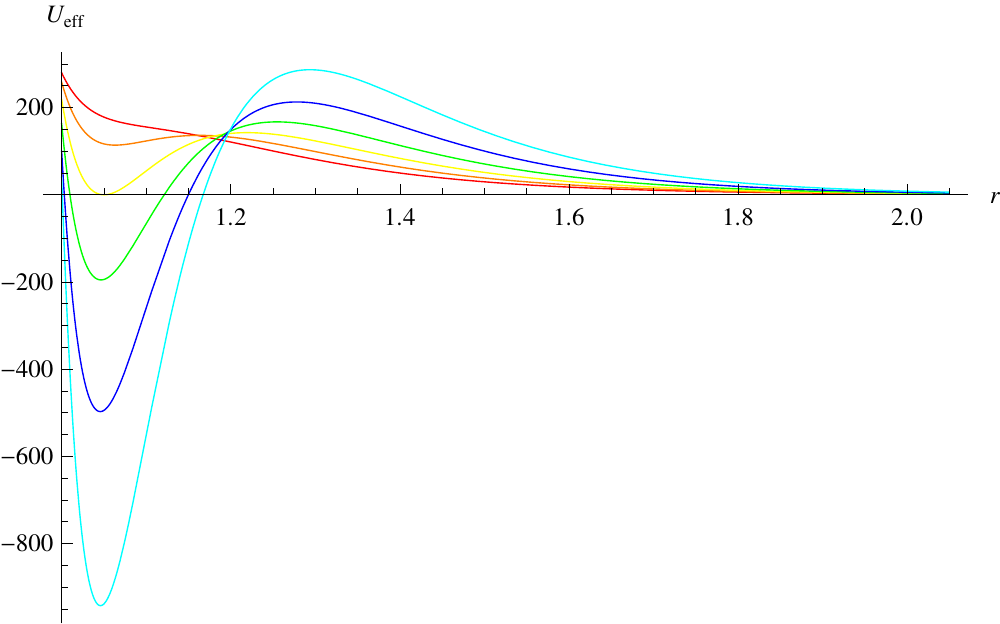}
\caption{The potential as a function of the radial coordinate for coupling constant $\kappa=-20$, mass and charge for the scalar field $q=0.6,\ \mu=1.63,\ \omega=1.28$. The ratio $(Q/M)^2$ is 0.852, 0.808, 0.760, 0.720, 0.679, 0.64 (red, orange, yellow, green, blue, cyan) respectively.} \label{fig2e}
\end{figure}

An interesting behaviour is observed if we fix the value of the derivative coupling $\kappa$. Then in Figure \ref{fig2c} we depict the effective potential as a function of $r$ for fixed values of $\omega, \mu, q, \kappa$ and we vary the values of change $Q$ and mass $M$ belonging to the first part of Table \ref{table1}. We see that as the ratio is increased eventually a trapping potential is formed. Note that this happening as the charge $Q$ is increased.
In Figure \ref{fig2d} we depict the effective potential  for values of $Q$ and $M$ which  they belong to the second part of the Table \ref{table1}. We observe that a trapping potential is formed only for values of $Q$ and $M$ which they give a ratio that it is far away from the extremal limit or the near extremal limit. Finally in Figure \ref{fig2e} we see the formation of a trapping potential for values of $Q$ and $M$ belonging to the third part of Table \ref{table1}. We observe that as the mass $M$ and charge $Q$ in increased the potential well gets deeper.

\section{Conclusions}

In this work we studied the superradiant instability of a charged scalar wave scattered off a Reissner-Nordstr\"om black hole. The scalar field except its minimal coupling to gravity it is also coupled kinematically to Einstein tensor. We find that the  Bekenstein's superrradiant condition $\omega < q \Phi$ is modified to $(r_H^2+\kappa \Phi^2) (\omega - q \Phi) < 0$, where $\kappa$ is the  derivative coupling constant.

In the  case of $\kappa$ positive (opposite sign than the canonical kinetic term) we find that for a wide range of parameters the superradiant condition is satisfied and in the same time a trapping potential is formed outside the horizon of a Reissner-Nordstr\"om black hole. As the strength of the coupling of the scalar field to curvature is increasing the depth of the potential is increasing indicating the amplification of the superradiant instability. The same amplification occurs for constant coupling as the mass and charge of the black hole is increasing.

In the case of negative $\kappa$, in spite of the Bekenstein's superradiant condition is violated, a trapping potential is formed and as before, the depth of the potential is increased as the absolute value of the derivative coupling is increased. However, if the derivative coupling is fixed, to have the formation of a trapping potential the mass and charge of the black hole should be large.

\acknowledgments

We thank V. Cardoso, C. Herdeiro and S. Hod for their valuable comments and remarks. The work of T.K. was funded by the FONDECYT Grant No. 3140261.


\end{document}